\documentclass[
 reprint,amsmath,amssymb,aps,
prb, longbibliography,
floatfix,
]{revtex4-2}
\usepackage{xcolor}  
\usepackage{placeins}
\usepackage{lineno}
\usepackage{soul}
\usepackage{graphicx,float}
\usepackage{amsmath,bm}
\usepackage{amssymb}
\usepackage{braket}
\usepackage{amsfonts}
\usepackage{dsfont}
\usepackage{comment,color}
\usepackage{lipsum}
\usepackage{hyperref}
\hypersetup{colorlinks=true,linkcolor=blue,anchorcolor=blue,citecolor=blue,filecolor=blue,urlcolor=blue,bookmarksnumbered=true,pdfview=FitB}

\begin{document}

\title{Complex spin dynamics induced metamagnetic phase transitions in Dirac semimetal EuAuBi}
\author{Lipika$^{1}$}
\author{Shobha Singh$^1$}
\author{Anyesh Saraswati$^2$}
\author{Vikas Chahar$^1$}
\author{Yan Sun$^{3,4}$}
\author{Pascal Manuel$^5$}
\author{Devashibhai Adroja$^{5,6}$}
\author{Walter Schnelle$^7$}
\author{Nitesh Kumar$^2$}
\author{Jhuma Sannigrahi$^8$}
\author{Kaustuv Manna$^{1,*}$}
\affiliation{$^1$Department of Physics, Indian Institute of Technology Delhi, Hauz Khas, New Delhi, India 110016}
\affiliation{$^2$S. N. Bose National Centre for Basic Sciences, Salt Lake, Kolkata-700 106, India}
\affiliation{$^3$Shenyang National Laboratory for Materials Science, Institute of Metal Research, Chinese Academy of Sciences, Shenyang, 110016, China}
\affiliation{$^4$School of Materials Science and Engineering, University of Science and Technology of China, Shenyang 110819, China}
\affiliation{$^5$ISIS Neutron and Muon Source, Science and Technology Facilities Council, Rutherford Appleton Laboratory,
Didcot OX11 0QX, United Kingdom}
\affiliation{$^6$Highly Correlated Matter Research Group, Physics Department, University of Johannesburg, P.O. Box 524,
Auckland Park 2006, South Africa}
\affiliation{$^7$ Max-Planck-Institute for Chemical Physics of Solids,Nöthnitzer Straße 40
01187 Dresden, Germany}
\affiliation{$^8$School of Physical Sciences, Indian Institute of Technology Goa, Farmagudi, Goa 403401, India}
\affiliation{$^*$kaustuvmanna@physics.iitd.ac.in}
\

\date{\today}
      
\begin{abstract}
We report a comprehensive investigation of the physical properties of the Dirac semimetal compound EuAuBi single crystals, using neutron diffraction, magnetization, electrical transport, and specific heat measurements. EuAuBi crystallizes in a hexagonal structure with space group \textit{P}6$_3$\textit{mc} (No. 186). First-principles calculations using density functional theory characterize it as a Dirac semimetal, with a notable band-crossing in proximity to the Fermi level (E$_F$) along the $\Gamma$-A direction. 
The crystal exhibits three distinct magnetic phases at 4 K (\textit{T$_{N1}$}), 3.5 K (\textit{T$_{N2}$}), and 2.8 K (\textit{T$_{N3}$}) as observed from magnetic and specific heat measurements. However, zero-field neutron diffraction resolves  only two magnetic phases: a commensurate antiferromagnetic phase and a canted antiferromagnetic phase. Field-dependent \textit{ac} and \textit{dc} magnetization measurements uncover field-induced non-trivial spin textures in the magnetic field range 1.5 to 3 T, manifested as a tilted plateau in the magnetization curves. The interplay between conduction carriers and these spin textures is further evidenced by unique features in the magnetic field-dependent longitudinal resistivity in the system. Finally, we present a comprehensive magnetic phase diagram of EuAuBi, highlighting diverse spin alignments present in the material. EuAuBi thus emerges as a rare material system in which both momentum-space and real-space Berry curvature effects may coexist, providing a unique opportunity to investigate their interplay.
\end{abstract}

\maketitle  
 \section{Introduction}
In recent years, quantum materials have emerged as a major thrust area in condensed matter physics. Understanding the complex physics of various topological states and their tunability opens new pathways for next-generation topo-spintronic applications \cite{FERT2019817, Chadov2010TunableCompounds}. In these quantum states, the presence of Berry curvature, either in real space or in momentum space is very crucial and is characterized by their topological stability. In momentum space (\textit{k}-space), the topological behaviour manifests in various materials like Nodal line \cite{Shao2020ElectronicSemimetals, Fang2016TopologicalSemimetals, Kim2018LargeSemimetal}, Dirac\cite{Young2012DiracDimensions, Armitage2018WeylSolids, doi:10.1126/science.1245085}, and Weyl semimetals \cite{Bernevig2022ProgressMaterials, Yan2017TopologicalSemimetals, Armitage2018WeylSolids, Soluyanov2015Type-IISemimetals}. 
The large Berry curvature arising from strong electron entanglement in these systems give rise to various exotic quantum effects \cite{Narang2021TheStructures, Manna2018FromDesign, doi:10.1126/sciadv.1501870}. Besides, when the symmetry-protected topological state exists close to the Fermi energy, it can strongly affect the physical properties leading to remarkable responses such as giant anomalous Hall conductivity in electrical transport or massive anomalous Nernst effect in thermal transport measurements. \cite{Nagaosa2010AnomalousEffect, Kumar2021TopologicalChemistry, Sun2018TopologicalCompounds, Guin2019AnomalousCo2MnGa, Nakatsuji2015LargeTemperatureb, Guin2019Zero-FielDCo3Sn2S2}. 
\newline
\begin{figure*}[t]
    \centering   
    \includegraphics[width=17cm,height=11cm]{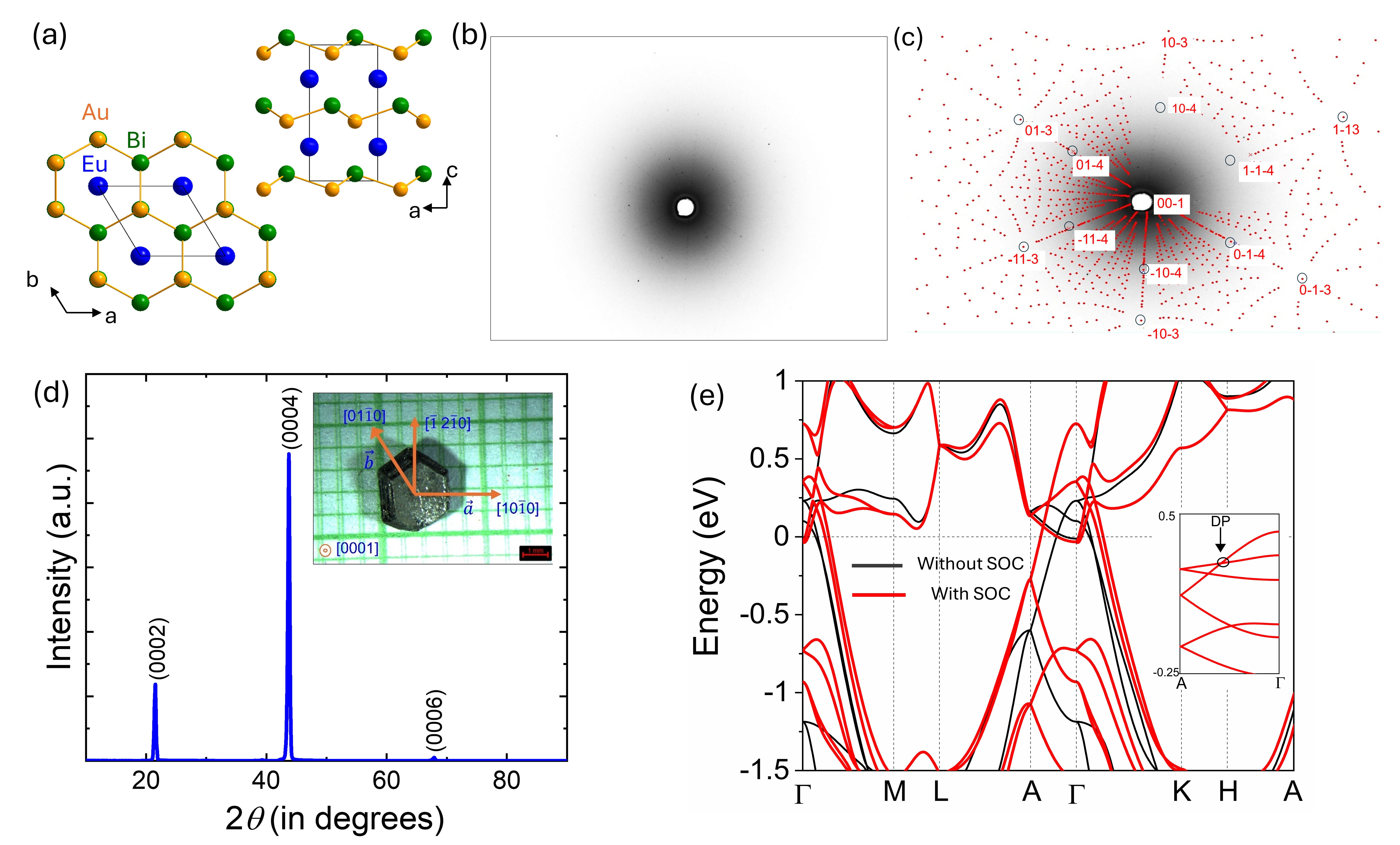}
    \caption{(a) Atomic arrangement of the EuAuBi crystal structure with the top and side view. Blue, Yellow and Green atoms represent the Eu, Au, and Bi atoms respectively. (b) Laue diffraction pattern of EuAuBi single crystal. (c) Raw Laue diffraction pattern is superimposed over a simulated pattern. (c) $\theta$ - 2$\theta$ XRD of the EuAuBi crystal, reflecting the peak orientations as indexed with the (000l) planes. The inset shows the grown EuAuBi crystal with the directions defined.  (d) The band structure of EuAuBi with and without SOC. The inset shows the Dirac point (DP) along the $\Gamma$-A direction.}
    \label{fig:my_label1}
\end{figure*}
In the spatial domain, real space Berry curvature effect is associated with the noncoplanar spin textures such as magnetic skyrmions \cite{Bogdanov2020PhysicalSkyrmions}. These topologically protected spin textures are robust against external perturbations. While the
Dzyaloshinskii–Moriya interaction (DMI) is a well-known
mechanism for stabilising skyrmions in noncentrosymmetric
systems, frustrated magnetic interactions or magnetic anisotropy can also stabilize
skyrmions in centrosymmetric lattices— thus broadening the
range of candidate materials \cite{Bogdanov2020PhysicalSkyrmions, Topologicalskyrmions, Felser2022TopologyCompounds, Kurumaji2019, Chakrabartty2022, Hou2017, Wang2016}. 
These systems hold significant promise for technological applications in memory devices, logic gates, and communication technology \cite{Fert2017MagneticApplications, Luo2021SkyrmionApplications}. 
\newline
 However, most studies so far have been focused on understanding the effect of either the momentum-space or the real-space Berry curvature on the physical properties of quantum materials. It is a rare coincidence where both effects coexist \cite{Vir2019AnisotropicPtSn}. Such systems offer unique opportunities and possibilities for a new degree of freedom to control or tune the \textit{k}-space quantum states with the real space elements or vice-versa \cite{Fujisawa2023WidelyCr1+Te2, Du2020BerryAntiferromagnets}. These smart materials can also be used for the development of multifunctional quantum devices, important for the next generation of topological spintronic applications. 
In this work, we investigate the topological semimetal EuAuBi that exhibits both real-space and momentum-space topological features. First-principles band structure calculations reveal symmetry-protected Dirac crossing near the Fermi energy along the $\Gamma$– A direction, pointing to a non-trivial momentum-space topology. However, in the present manuscript, we primarily focus on understanding the real-space topological effect in EuAuBi. We conducted detailed \textit{ac} and \textit{dc} magnetization, zero-field neutron diffraction, specific heat, and electrical transport measurements to investigate the physical properties of EuAuBi single crystals.
 While earlier studies have explored high critical magnetic-field superconductivity in EuAuBi \cite{Takahashi2023}, our work primarily focuses on the intricate spin arrangements and field-induced magnetic textures present above 2 K in EuAuBi single crystals. The distinct first-order phase transition in field and temperature dependent magnetization, as well as specific heat, together with frequency dependent peak shift in \textit{ac} susceptibility, indicates the possible presence of non-trivial spin texture in the EuAuBi system. Based on the detailed experimental analysis, we have constructed a comprehensive magnetic phase diagram for EuAuBi.

\begin{figure*}[t]
    \centering
    \includegraphics[width=17cm]{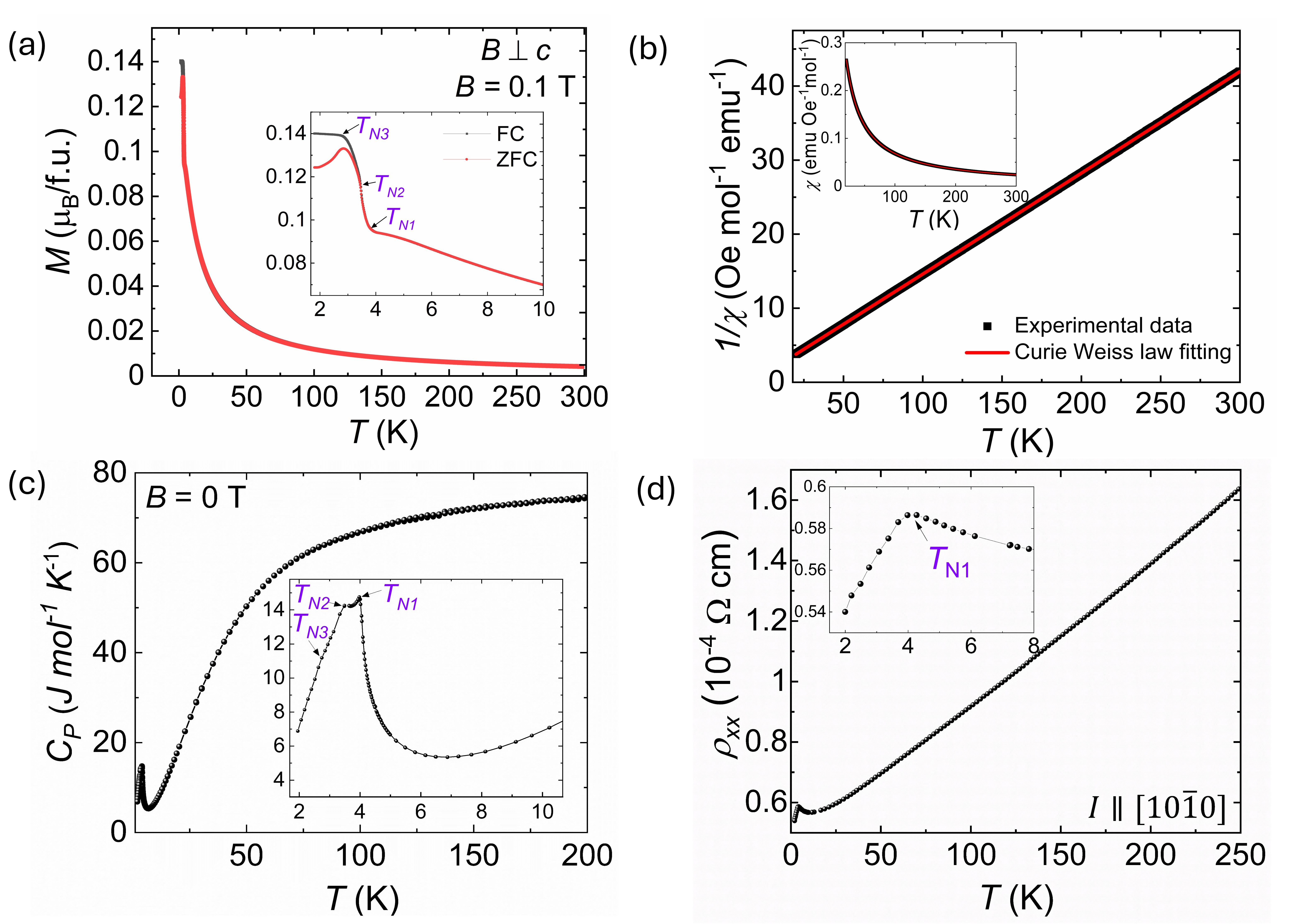}
    \caption {(a) Temperature dependent zero-field-cooling (ZFC) and field-cooling (FC) magnetization \textit{M}(\textit{T}) for \textit{B} = 0.1 T applied field in the \textit{ab} plane of crystal (\textit{B} $\perp$ \textit{c}). Inset shows a zoomed-in picture from 2 - 10 K, which indicates three transitions present in the system at \textit{$T_{N1}$}, \textit{$T_{N2}$}, and \textit{$T_{N3}$} (b) Temperature-dependent inverse susceptibility (1/$\chi$) in the paramagnetic regime (overlaps for \textit{B} $\parallel$ \textit{c} and \textit{B} $\perp$ \textit{c}) follows the extended Curie-Weiss law. The inset shows the temperature dependence of susceptibility ($\chi$(\textit{T})). (c) Temperature-dependent specific heat (\textit{C$_p$}(\textit{T})) at \textit{B} = 0 T. Inset shows zoomed-in picture from 2 - 10 K showing three transition temperatures similar to \textit{M}(\textit{T}). (d) Temperature dependent longitudinal resistivity (\textit{$\rho_{xx}$}(\textit{T})) for \textit{B} = 0 T and current applied in the \textit{ab} plane of the crystal (\textit{I} $\parallel$ [10$\bar{1}$0]). A zoomed-in view of \textit{$\rho_{xx}$}(\textit{T}) around the transition (\textit{$T_{N1}$}) is shown in the inset.}
    \label{fig:my_label2}
\end{figure*}
\section{Experimental Details}
Single crystals of EuAuBi were grown using the flux method with Bismuth as the flux \cite{Takahashi2023}. Elements Eu (Alfa Aeser, 99.99 \%), Au (Alfa Aeser, 99.99 \%), and Bi (99.99 \%) were weighed in a molar ratio of 1:1:10. Prior to use, bismuth was purified using the zone refinement technique by repeated melting in an inert atmosphere and removing the outer-side impurity layer. As Europium is  air sensitive, it was handled inside an argon-filled glove box. The weighed elements were placed in an alumina crucible, and vacuum sealed in a quartz tube. The ampoule was heated to 1000 °C at the rate of 100 °C/h, held at that temperature for 12 hours, and then slowly cooled to 400 °C at the rate of 5 °C/h . At 400°C, the ampoule was centrifuged to separate crystals from excess flux. The silverish-black, plate-like hexagonal-shaped crystals of approximately  2 - 3 mm were obtained (picture of one crystal is shown in fig 1c inset). EuAuBi crystals were also successfully synthesized using lead flux, where centrifuge temperature was kept higher to avoid the formation of any binary intermetallic impurity. The physical properties observed in both Bi- and Pb-flux-grown EuAuBi crystals were found to be qualitatively similar.
\begin{figure*}[t]
 \centering   
    \includegraphics[width=17 cm]{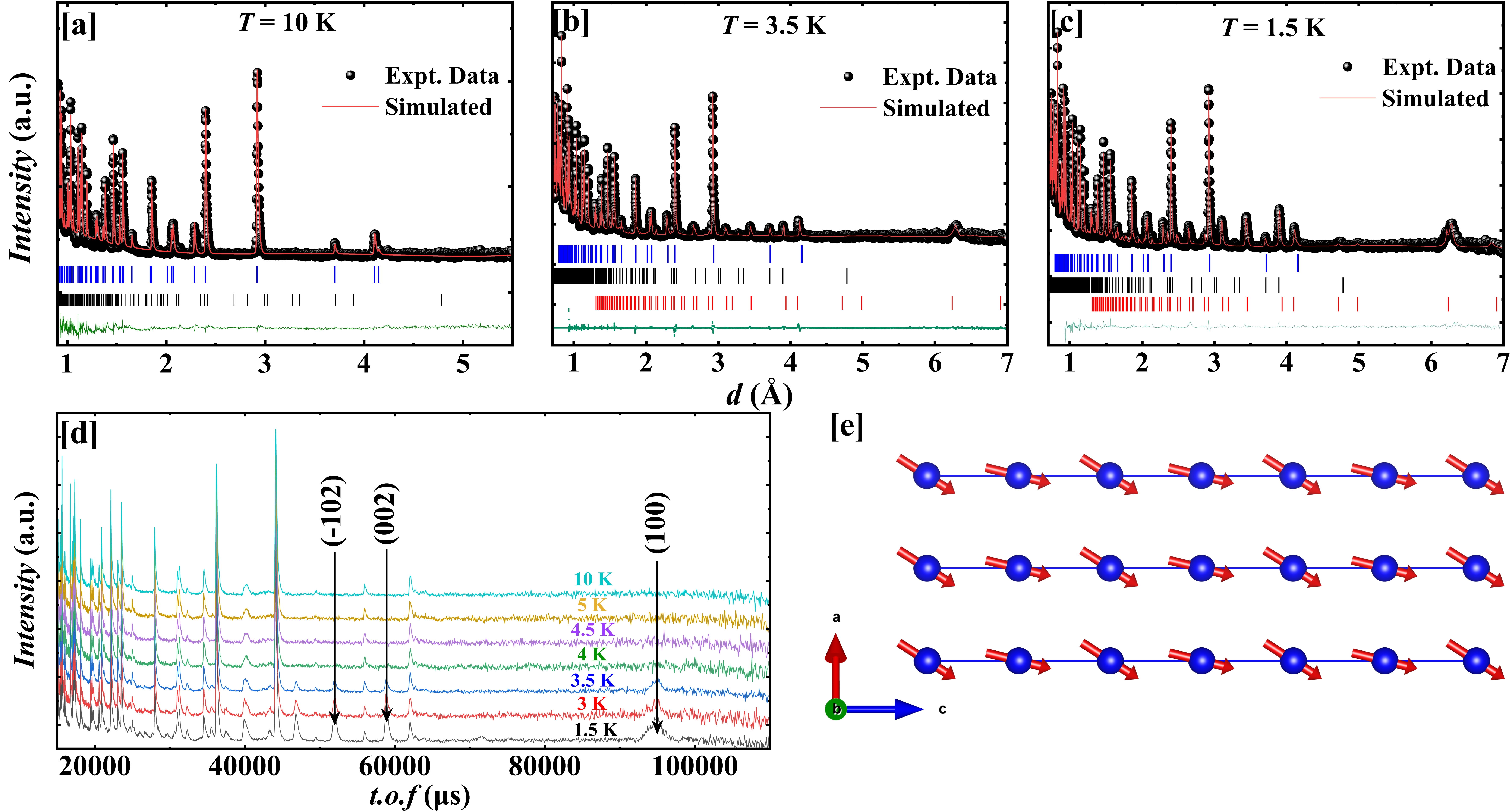}
    \caption{PND patterns of EuAuBi collected at different temperatures along with Rietveld refinements: (a) 10 K (paramagnetic region), (b) 3.5 K, and (c) 1.5 K (below the long-range ordering temperature). The blue and black vertical ticks indicate the Bragg peak positions corresponding to the main EuAuBi phase and a trace amount of the binary AuEu impurity phase, respectively. The red ticks in (b) and (c) mark the positions of the magnetic Bragg reflections. The green line below the Bragg markers represents the difference between the experimental and calculated profiles. Panel (d) displays the temperature-dependent PND patterns in the 1.5–10 K range, illustrating the evolution of magnetic Bragg peaks below the ordering temperature. Panel (e) presents a perspective view of the refined Eu spin configuration at 1.5 K.}
    \label{fig:my_label3}
\end{figure*}
\newline The elemental composition of the crystal was investigated using Energy dispersive X-ray spectroscopy (Hitachi- High Technol TM 3000) with 15 keV energy. The elemental composition of Eu, Au, and Bi were found approximately in the stoichiometric ratio of 1:1:1 within the experimental error. The crystal orientation was confirmed using Laue diffraction. The $\theta$ - 2$\theta$ X-ray diffraction was performed using HR-XRD (PANalytical Empyrean X-Ray diffractometer) with a Cu K$\alpha$ (\textit{$\lambda$} = 1.5406 \AA) monochromatic source at room temperature. The crystal structure and the microscopic spin alignment were investigated using neutron diffraction at the ISIS beamline facility. The in-plane and out-of-plane \textit{dc} and \textit{ac} magnetic measurements were carried out using the commercial 7 T Quantum design Magnetic Property measurement system (MPMS-3) and 14 T cryogen-free measurement system from Cryogenic Limited. Electrical transport and specific heat measurements were performed using the commercial Quantum Design Physical Property Measurement System (PPMS). For resistivity measurements, a standard linear four-probe geometry was used, using 25-micron platinum wire and silver paint for electrical contacts. 
\newline
First-principle-based electronic structure calculations were performed based on the code of Vienna ab initio simulation package (VASP)  with the projected augmented wave potential \cite{PhysRevB.54.11169}. The exchange and correlation energies were considered in the generalized gradient approximation (GGA), following Perdew–Burke–Ernzerhof (PBE) parametrization scheme \cite{PhysRevLett.77.3865}.
\newline
Neutron diffraction experiment was performed on a powder sample of EuAuBi at the ISIS Facility (UK) where high-resolution data were collected at the WISH diffractometer \cite{DrJhumaSannigrahi2018MagneticCrystals}. The powder sample of EuAuBi (weighing about 4 g) was placed in a thin hollow-shaped cylinder in order to mitigate the effect of the strong neutron absorption of Eu nuclei as much as possible. The neutron data were collected on heating at seven different temperatures (below and above the long-range ordering temperatures) in the range of 1.5 to 10 K. Rietveld refinements were carried out on the collected neutron powder diffraction data using the FULLPROF and JANA2020 program \cite{RODRIGUEZCARVAJAL199355}.
\section{Results and Discussion}
EuAuBi crystallizes in a hexagonal crystal structure with the space group $\textit{P}$$6_{3}$$\textit{mc}$ (186) \cite{Merlo1990RMXBi}. Its crystal structure features a honeycomb-like arrangement of Bi and Au atoms in the \textit{ab}-plane, while Eu atoms occupy the interstitial sites along the \textit{c}-direction between the two honeycomb layers as shown in fig 1a.
Laue diffraction measurements (fig 1b), depicts
distinct and defined spots indicating good crystalline quality without twinning or domains. The Laue indexing (fig 1c) reveals that the naturally grown single crystals are \textit{c}-oriented and the surface exposed to the X-rays is parallel to the \textit{ab} plane (inset of fig 1d). This is also evident from the $\theta$ - 2$\theta$ XRD as shown in fig 1d \cite{10.1063/1.4812323}. 
\newline
In order to understand the physical properties of the system, we conducted detailed temperature- and field-dependent \textit{ac} and \textit{dc} magnetization and electrical transport measurements. The experiments were perfomed by applying magnetic field along the \textit{c}-direction (\textit{B} $\parallel$ \textit{c}) and along various in-plane directions (\textit{B} $\perp$ \textit{c}), i.e., \textit{B} $\parallel$ [10$\bar{1}$0], \textit{B} $\parallel$ [$\bar{1}$2$\bar{1}$0], \textit{B} $\parallel$ 01$\bar{1}$0], etc. When the magnetic field was applied in any direction parallel to the crystal's \textit{ab} plane, the results were comparable, suggesting that the crystal is isotropic in the \textit{ab} plane. Therefore, in the further text, we mentioned \textit{B} $\perp$ \textit{c} for the magnetic field applied in all in-plane field orientations.
The temperature-dependent zero-field cooling (ZFC) and field cooling (FC)  magnetization curves of EuAuBi crystal for a magnetic field of 0.1 T applied along \textit{B} $\perp$ \textit{c} are shown in Fig 2a. Three distinct transitions are observed at approximately 4 K (\textit{$T_{N1}$}), 3.5 K (\textit{$T_{N2}$}), and 2.8 K (\textit{$T_{N3}$}) (Fig 2a inset), suggesting successive changes in the spin alignment of the system at these temperatures. These transitions are consistent with the earlier report on EuAuBi { \cite{Takahashi2023}}. A bifurcation occurs in the FC - ZFC curves upon cooling below \textit{$T_{N2}$} as shown in fig 2a inset. However, above $\sim$ 4 K, ZFC and FC curves superimpose and exhibit typical paramagnetic behaviour. To gain deeper insights into these transitions, we analyzed the temperature-dependent inverse \textit{dc} susceptibility (\textit{$\chi^{-1}$}), specific heat (\textit{C$_p$}), and longitudinal resistivity (\textit{$\rho_{xx}$}).
\begin{figure*}[t]
    \centering   
    \includegraphics[width=17cm]{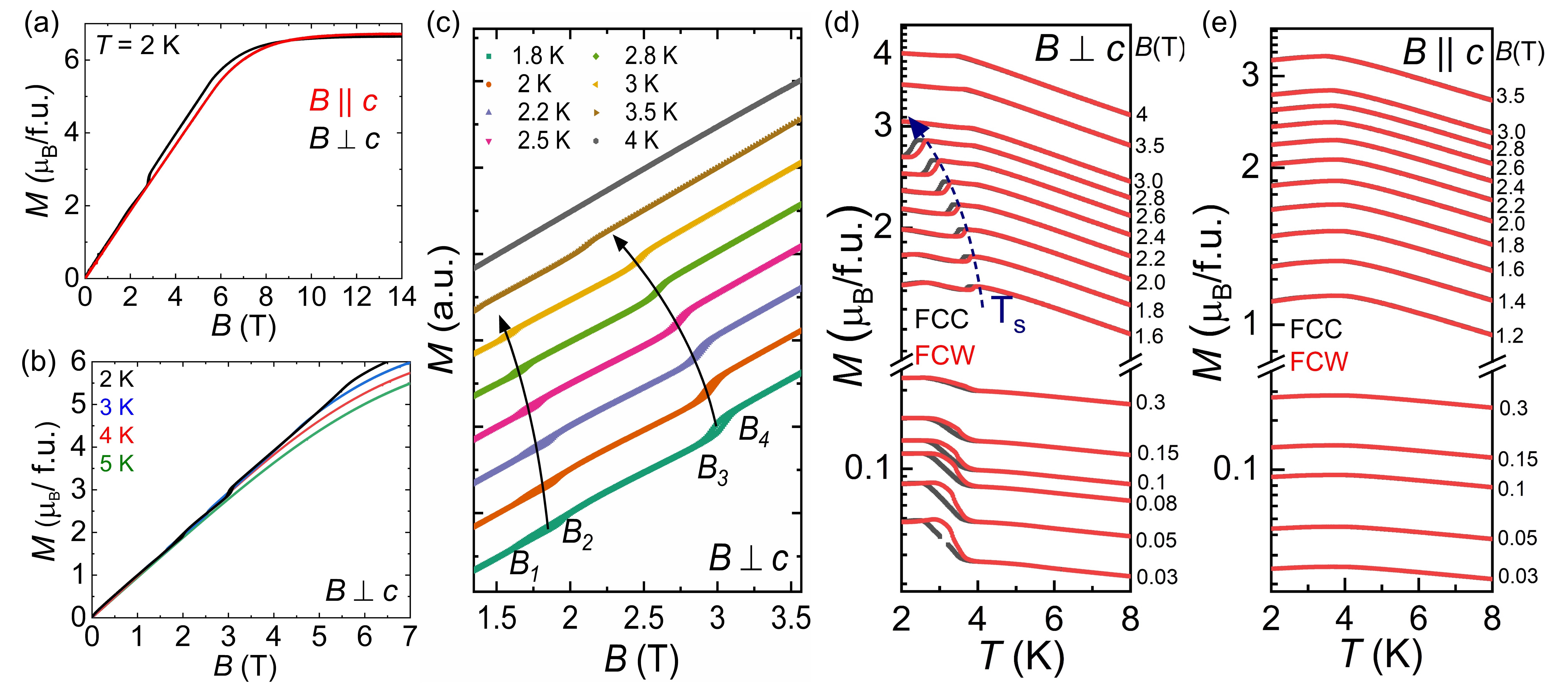}
    \caption
{(a) Isothermal field-dependent magnetization from 0 to 14 T \textit{dc} magnetic field at \textit{T} = 2 K for \textit{B} $\parallel$ \textit{c} and \textit{B} $\perp$ \textit{c} configuration. (b) Isothermal field-dependent magnetization at different temperatures with \textit{dc} field from 0 - 7 T \textit{dc} field for \textit{B} $\perp$ \textit{c} configuration. (c) Zoomed-in view of (b), from 1.5 to 3.5 T, revealing the metamagnetic transitions. \textit{B$_1$} and \textit{B$_2$} are marked as the starting and ending of the hysteresis around 2 T, \textit{B$_3$} and \textit{B$_4$} are marked as the starting and ending of the hysteresis around 3 T. The arrow highlights the shift in hysteresis towards lower magnetic fields with increasing temperature. Temperature-dependent magnetization at different magnetic fields for field-cooled cooling (FCC) and field-cooled warming (FCW) curves shown in black and red colors, respectively for (d) \textit{B} $\perp$ \textit{c} and (e) \textit{B} $\parallel$ \textit{c}. \textit{T$_S$} denotes the middle of the thermal hysteresis for \textit{B} $\geq$ 1.6 T and the arrow shows the shifting of hysteresis towards lower temperature as \textit{B} increases.}
    \label{fig:my_label4}
\end{figure*}
\begin{figure*}[t]
{
    \centering   
    \includegraphics[width=18cm]{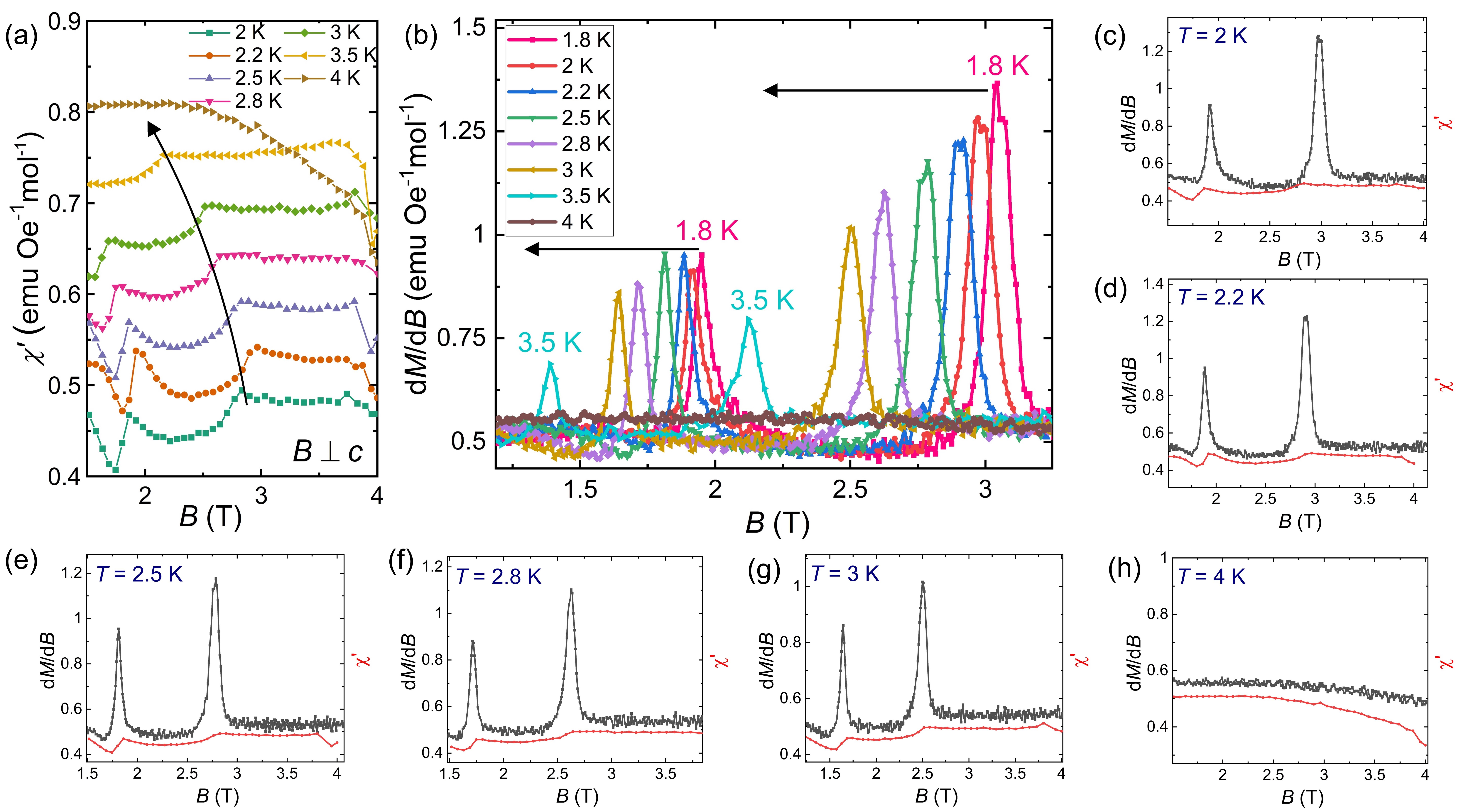}
    \caption{(a) Field-dependent real part of \textit{ac} susceptibility ($\chi$$^\prime$(\textit{B})) for \textit{B} $\perp$ \textit{c} at different temperatures. An \textit{ac} field of 9 Oe with frequency 93 Hz, is superimposed over the varying \textit{dc} field.(b) Field dependent d\textit{M}/d\textit{B} at different temperatures (as derived from \textit{M}(\textit{B}) in fig 4c). Comparison of d\textit{M}/d\textit{B} and $\chi$$^\prime$ at (c) \textit{T} = 2 K, (d) \textit{T} = 2.2 K, (e) \textit{T} = 2.5 K, (f) \textit{T} = 2.8 K, (g) \textit{T} = 3 K, (h) \textit{T} = 4 K. }
    \label{fig:my_label5}
    }
\end{figure*} 

The temperature-dependent \textit{$\chi^{-1}$}(\textit{T}) above 10 K is shown in Fig 2b. It follows the extended Curie-Weiss law $\chi$ = C/(\textit{T}-$\theta$)+$\chi_o$, where C is the Curie constant, $\theta$ is paramagnetic Curie-Weiss temperature, and $\chi_o$ is temperature independent susceptibility. We found $\theta$ = -8.02 K, suggesting predominant antiferromagnetic interaction in the system. The effective magnetic moment is found to be $\approx$ 7.70 $\mu_B$, which is close to the theoretical moment g$\sqrt{S(S+1)}$ $\mu_B$ = 7.94 $\mu_B$ per Eu ion \cite{Takahashi2023}.
Fig 2c shows the specific heat \textit{C$_p$}(\textit{T}) of EuAuBi in the temperature range 1.8 - 200 K under zero applied magnetic field. The specific heat saturates to 74.15 Jmol$^{-1}$K$^{-1}$ near 200 K, which matches with the classical value obtained from the Dulong-Petit law: \textit{C$_p$} = 3NR i.e. 74.82 Jmol$^{-1}$K$^{-1}$, where N = 3 for EuAuBi. As shown in the inset of fig 2c, a clear \textit{$\lambda$}-shaped peak around \textit{$T_{N1}$} along with two additional kinks at \textit{$T_{N2}$}, and \textit{$T_{N3}$} are observed, consistent with the anomalies seen in the magnetization measurements \cite{Takahashi2023}. This strongly suggests notable changes in the spin configuration of EuAuBi at \textit{$T_{N1}$}, \textit{$T_{N2}$}, and \textit{$T_{N3}$} as system is cooled below 4 K.
The temperature dependent resistivity \textit{$\rho_{xx}$}(\textit{T}) in zero magnetic field in the temperature range 2 K to 250 K with \textit{I} $\parallel$ [2-1-10], is shown in Fig 2d. Above 10 K, \textit{$\rho_{xx}$} increases linearly with temperature, indicating the metallic behaviour of the crystals, dominated by phononic contribution. However, below 10 K, \textit{$\rho_{xx}$} shows an upturn with decreasing temperature and exhibits a broad hump around 4 K, corresponding to Néel temperature \textit{$T_{N1}$} of the system.  However, the other two transitions (\textit{$T_{N2}$} and \textit{$T_{N3}$}) are not observed in the zero-field \textit{$\rho_{xx}$}(\textit{T}) curve. This suggests that either these two transitions are smeared off behind the broad hump or have a negligible impact on the electical transport. However, in contrast to the previous report \cite{Takahashi2023}, we have not observed any superconducting transitions in our EuAuBi single crystals. 

\subsection{Powder Neutron Diffraction}
To gain insights into the spin ordering in EuAuBi, zero-field powder neutron diffraction (PND) measurements were performed at selected temperatures between 10~K and 1.5~K. The powder sample used for the PND measurements was synthesized via a standard solid-state reaction route. The diffraction pattern at 10~K, corresponding to the paramagnetic state, is well described by the reported hexagonal structure (space group $P6_{3}mc$, No.~186) with refined lattice parameters $a = b = 4.772(7)$~\AA~and $c = 8.243(4)$$~\AA$ \cite{Merlo1990RMXBi}. In addition to the main reflections of EuAuBi, weak extra peaks corresponding to trace amounts of the AuEu impurity phase (space group $Pnma$) were also observed (see Fig. 3a-c). Upon cooling to 1.5~K, several additional Bragg peaks emerge as shown in Fig 3d, clearly indicating the onset of long-range magnetic order. The refined magnetic structure of EuAuBi at low temperature corresponds to a commensurate canted antiferromagnetic state characterized by the propagation vector $\mathbf{k} = (1/3, 0, 0)$. Indexing of the magnetic Bragg peaks yields a propagation vector very close to $(1/3, 0, 0)$. Within the resolution of the WISH dataset and the strong neutron absorption of Eu, no evidence for slight incommensurability was observed, and the commensurate value $\mathbf{k} = (1/3, 0, 0)$ provides the most stable and physically meaningful refinement. The Eu moments mostly lie within the $ac$-plane, forming a gently non-collinear arrangement (Fig. 3e). A slight out-of-plane canting introduces a periodic alternation in the spin orientation between neighboring Eu layers, resulting in a modulated but symmetry-consistent magnetic configuration.  
\newline
Magnetic symmetry analysis was carried out in \textsc{Jana2020} using both the Shubnikov magnetic space-group and magnetic superspace-group formalisms. The analysis identifies two magnetically inequivalent Eu positions generated by the propagation vector $\mathbf{k} = (1/3, 0, 0)$ at (0, 0, 0.25) and (0, 0, 0.75). Two possible magnetic irreducible representations (IRs) were tested: $m\mathrm{SM1}$ [$Cm1^{\prime}(a0g)0s$] and $m\mathrm{SM2}$ [$Cm1^{\prime}(a0g)ss$], with $m\mathrm{SM2}$ providing the best agreement with the observed magnetic intensities. The refinement confirms a canted, non-collinear magnetic structure with $\mathbf{k} = (1/3, 0, 0)$. The Eu moments mostly lie within the ac plane, but exhibit a small out-of-plane canting that generates a three-site periodic modulation in their orientations. This modulation arises from the differing phase relationships between the in-plane and out-of-plane components of the moments, causing each spin to tilt slightly away from perfect collinearity. The refined envelope magnetic moments are 5.94~$\mu_{\mathrm{B}}$ for Eu1 and 5.64~$\mu_{\mathrm{B}}$ for Eu2, values that are close to the expected moment for Eu$^{2+}$ ($S = 7/2$), confirming well-localized 4$f$ moments. This canted commensurate arrangement explains the enhanced magnetic Bragg intensities observed at low temperature and reflects the competition between in-plane exchange interactions and anisotropy within the Eu sublattice. Magnetization and specific-heat measurements reveal additional anomalies near 3.5~K and 2.8~K, suggesting intermediate rearrangements of the spin configuration. However, due to the strong neutron absorption of Eu and the limited scattering intensity from the powder samples, the neutron data could not be refined with sufficient accuracy to resolve these intermediate magnetic phases unambiguously. In particular, attempts to model additional reflections in the 3-4 K range did not yield stable or unique solutions.

\begin{figure*}[t]
{
    \centering   
    \includegraphics[width=17cm]{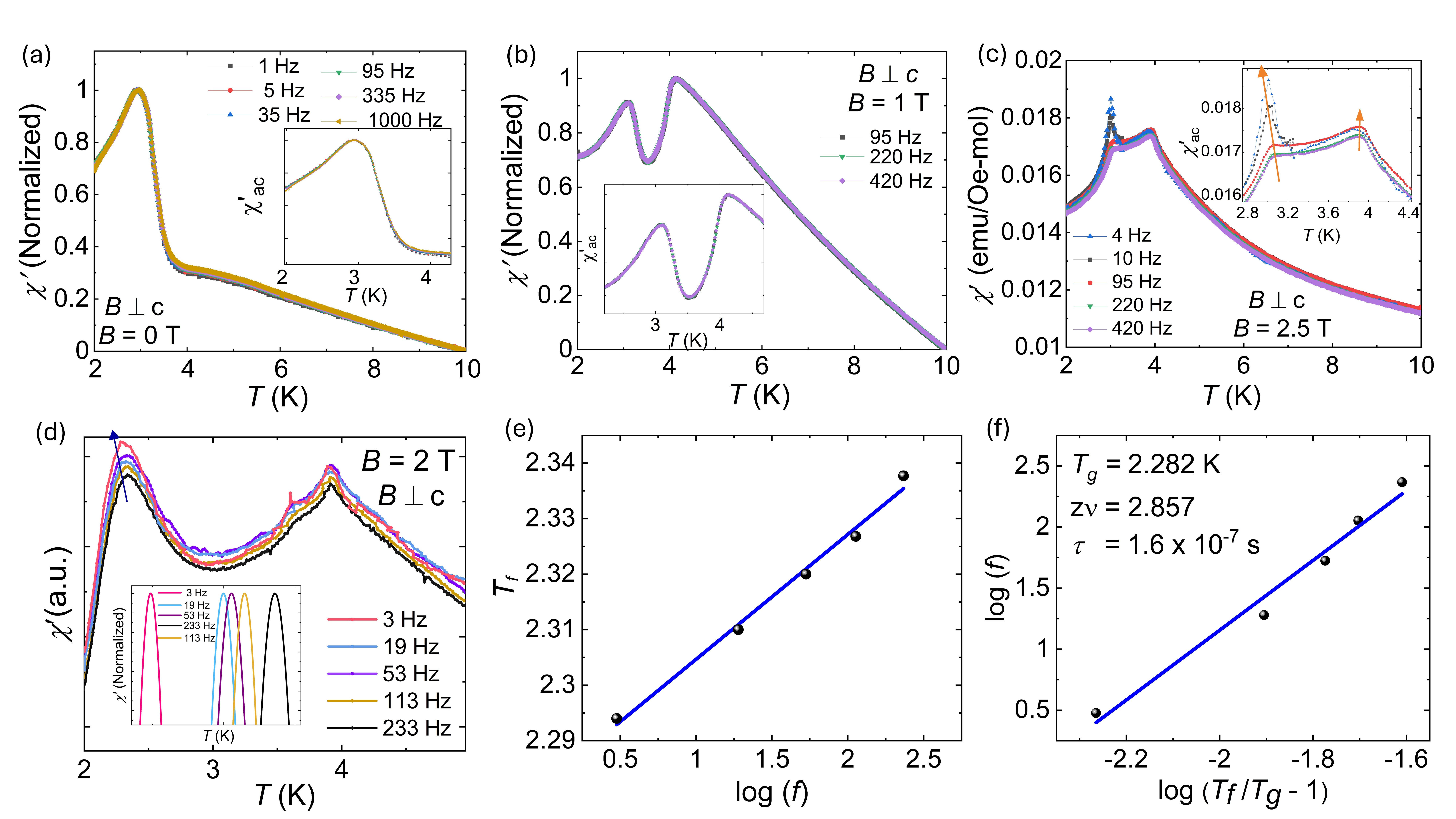}
    \caption{Temperature-dependent real part of \textit{ac} susceptibility, $\chi$$\prime$ at different frequencies for \textit{B} $\perp$ \textit{c} at applied \textit{dc} fileds, (a) \textit{B} = 0 T, (b) \textit{B} = 1 T, (c) \textit{B} = 2.5 T, (d) \textit{B} = 2 T. (e) Plot of spin-freezing temperature \textit{T$_f$} vs log \textit{f} obtained at \textit{dc} field of 2 T. (f) Plot of log(\textit{f}) vs log(\textit{T$_f$}/\textit{T$_g$} - 1), fitted using critical slowing down model and obtained parameters are mentioned in the figure. For \textit{ac} $\chi$$\prime$ measurements, 9 Oe of \textit{ac} field amplitude with 93 Hz frequency is superimposed over a constant \textit{dc} field. Inset of figs 6a-6c show the zoomed-in \textit{$\chi$$\prime$}(\textit{$T$}) and inset of fig 6d shows the zoomed-in normalized and polynomial fitted curves of $\chi$$\prime$ where frequency-dependent shift is clearly visible. }
    \label{fig:my_label6}
    }
\end{figure*} 
\subsection{ Field- and Temperature- dependent \textit{dc} Magnetization, \textit{M}(\textit{B}) and \textit{M}(\textit{T})}
We performed isothermal field-dependent magnetization measurements \textit{M}(\textit{B}) to investigate the effect of magnetic field on different spin alignments present in the system. Fig 4a shows \textit{M}(\textit{B}) results recorded at \textit{T} = 2 K for  \textit{B} $\perp$ \textit{c} and \textit{B} $\parallel$ \textit{c} in the range 0 - 14 T. For both field directions, the magnetization initially increases linearly with the field, as expected for the antiferromagnetic system. However, for \textit{B} $\perp$ \textit{c}, a metamagnetic transition is observed around 3 T. Also in this configuration, \textit{M} reaches saturation at a lower field compared to \textit{B} $\parallel$ \textit{c}, indicating that the \textit{ab} plane is the easy plane, while the \textit{c} axis is the hard axis of the crystal \cite{Takahashi2023}. Since, no metamagnetic transition is observed for \textit{B} $\parallel$ \textit{c} at any temperature, further \textit{M}(\textit{B}) results are discussed for the \textit{B} $\perp$ \textit{c} configuration. Such metamagnetic transitions have been previously observed in many Eu-based ternary systems such as EuCuBi, EuCuAs, EuAuAs, etc {\cite{PhysRevB.102.174425, PhysRevB.108.115126, Malick2022ElectronicCrystal, Roychowdhury2023InterplayEuCuAs}.
\begin{figure*}[t]
{
    \centering   
    \includegraphics[width=17cm]{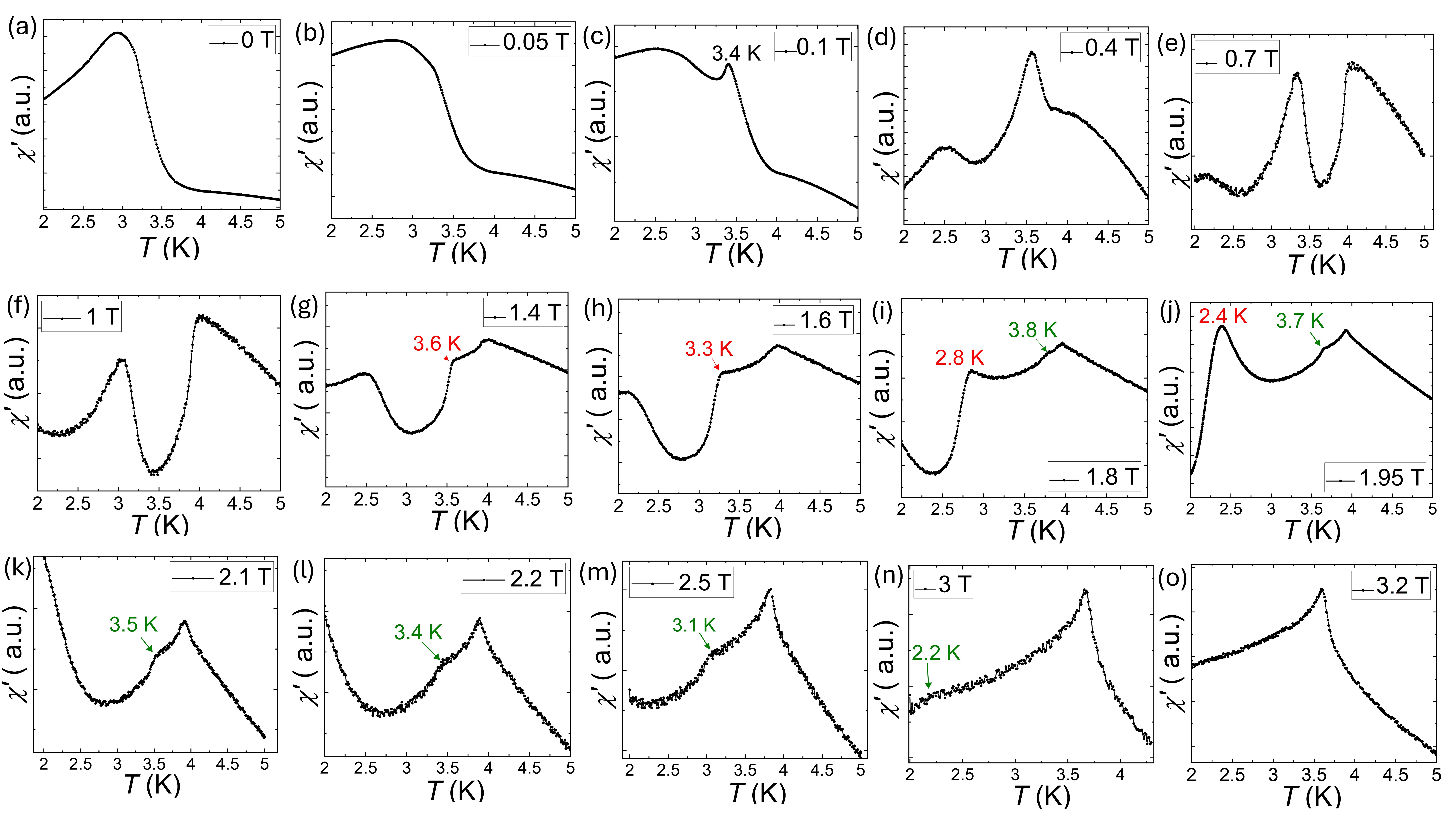 }
    \caption{Distinct features are observed in the temperature-dependent real part of \textit{ac} susceptibility for \textit{B} $\perp$ \textit{c}, \textit{$\chi$$^\prime$}(\textit{T}) in the presence of different \textit{dc} fields - (a) \textit{B} = 0 T, (b) \textit{B} = 0.05 T, (c) \textit{B} = 0.1 T, (d) \textit{B} = 0.4 T, (e) \textit{B} = 0.7 T, (f) \textit{B} = 1 T, (g) \textit{B} = 1.4 T, (h) \textit{B} = 1.6 T, (i) \textit{B} = 1.8 T, (j) \textit{B} = 1.95 T, (k) \textit{B} = 2.1 T, (l) \textit{B} = 2.2 T, (m) \textit{B} = 2.5 T, (n) \textit{B} = 3 T, (o) \textit{B} = 3.2 T. Here \textit{ac} field amplitude of 9 Oe and 93 Hz frequency is superimposed over different \textit{dc} magnetic fields.  The marked red and green temperatures from fig 7g to 7n show the possible first and second boundaries of the field-induced non-trivial phase in EuAuBi.}
    \label{fig:my_label7}
    }
\end{figure*} 
\newline
Fig 4b shows \textit{M}(\textit{B}) for \textit{B} $\perp$ \textit{c} at various temperatures, with the field swept from 0 T to 7 T and then back to 0 T. Hysteresis is observed in certain field ranges for \textit{T} $\textless$ 4 K. As shown in the enlarged view (Fig 4c), two distinct hysteresis are present between \textit{B$_1$} - \textit{B$_2$} and \textit{B$_3$} - \textit{B$_4$}, which shift towards a lower magnetic field with increasing temperature. This suggests that field-induced effects or competition between the different spin alignments might be prevailing in the system. A similar feature is reported in \cite{Takahashi2023}, where it is attributed to a spin-flop transition. Our findings further support this interpretation, indicating that the observed transitions in EuAuBi likely arise from spin reorientation under an applied field.
\newline
\begin{figure*}[t]
{
    \centering   
    \includegraphics[width=18cm]{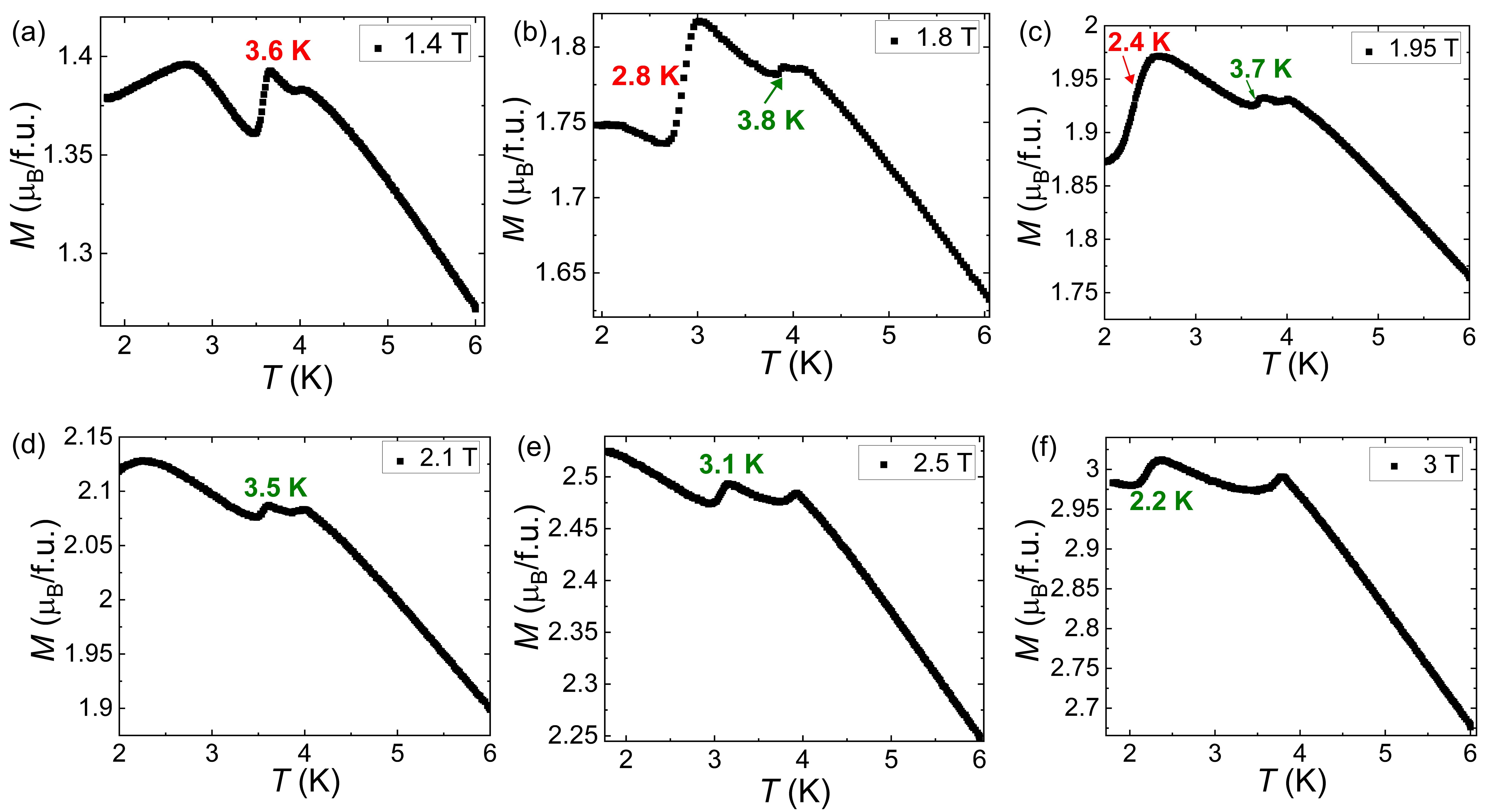}
    \caption{ The Zero field cooling (ZFC) temperature-dependent Magnetization \textit{M}(\textit{T}) in the presence of \textit{B} $\perp$ \textit{c} at (a) \textit{B} = 1.4 T, (b) \textit{B} = 1.8 T, (c) \textit{B} = 1.95 T, (d) \textit{B} = 2.1 T, (e) \textit{B} = 2.5 T, (f) \textit{B} = 3 T. The marked red and green temperatures show the possible first and second boundary of non-trivial field-induced magnetic phase in EuAuBi.}
    \label{fig:my_label8}
    }
\end{figure*} 
To further investigate the magnetic anomalies, we performed temperature-dependent magnetization \textit{M}(\textit{T}) under different applied magnetic fields. Fig 4d and 4e show the field-cooled cooling (FCC) and field-cooled warming (FCW) magnetization for \textit{B} $\perp$ c and \textit{B} $\parallel$ \textit{c}  respectively. These measurements highlight the presence of thermal hystersis in the magnetization response. For \textit{B} $\perp$ \textit{c}, at low magnetic field (up to 0.3 T), the thermal hysteresis between FCC and FCW curve remains nearly temperature independent in the range \textit{$T_{N1}$} - \textit{$T_{N3}$}. The observed thermal hysteresis is consistent with first-order–like behaviour, which has been reported in some systems with incommensurate or competing spin orders \cite{Das2014ObservationEuMnO3,Chen2013,GeerkenGriessen1982}. 
Further, as the magnetic field increases, the FCC and FCW curves gradually overlap, indicating a suppression of thermal hysteresis possibly due to commensurate magnetic structure.} Interestingly, beyond \textit{B} = 1.5 T, a new hysteresis loop emerges near \textit{$T_{N1}$} which shifts towards lower temperatures with increasing field as indicated as \textit{T$_S$}. This suggests the presence of another field-induced first-order phase transition in the system which shifts with increase in field. In contrast, for \textit{B} $\parallel$ c, no thermal hysteresis is observed across the measured field range, as shown in Fig 4e. Notably, for both field orientations, Néel temperature \textit{$T_{N1}$} for the commensurate antiferromagnetic transition shifts towards lower temperature with increasing magnetic field, a common feature of antiferromagnetic systems \cite{BlundellStephen2001, Lidiard1954}.

\subsection{Field dependent \textit{$\chi^\prime$} and d\textit{M}/d\textit{B}} 
To further investigate the magnetic transitions, we performed the field-dependent \textit{ac} susceptibility measurements in the field range 1.5 to 4 T (the real part of the \textit{ac} susceptibility signal \textit{$\chi$$^\prime$}(\textit{$B$}) is shown in fig. 5a). These measurements were carried out using an \textit{ac} excitation field of  9 Oe and frequency 93 Hz, superimposed on varying \textit{dc} magnetic fields. 
\textit{$\chi$$^\prime$}(\textit{$B$}) exhibits peaks at two distinct magnetic fields, forming a plateau-like region between them. Such features have been reported in MnSi and related systems, where they are commonly associated with the boundaries of skyrmion lattice phase \cite{Bauer2012, Bauer2010, Butykai2017, Bannenberg2018MagnetizationMn1-xFexSi, Qian2016PhaseOSeO3, Bannenberg2016, Gregory1992}. Here also, these peaks likely represent the boundaries of the field-induced non-trivial spin states, where the plateau region shows the robustness against small perturbations created by the applied \textit{ac} field. The plateau gradually shifts toward lower magnetic fields as temperature increases, and it vanishes around 4 K as the system transitions into a paramagnetic state.
To correlate this behavior with \textit{dc} magnetization, fig 5b presents the derivative of magnetization with respect to the applied field, d\textit{M}/d\textit{B}, extracted from \textit{M}(\textit{B}) (Fig 4c). The d\textit{M}/d\textit{B} plots exhibit sharp peaks, corresponding to step-like changes in \textit{M}(\textit{B}), which also shift to lower fields with increasing temperature. The intensity of these peaks decreases with temperature, indicating a suppression of the observed phase until it vanishes near 4 K.
Fig 5c–5h compare d\textit{M}/d\textit{B} from \textit{dc} magnetization with the real part of \textit{ac} susceptibility \textit{$\chi$$^\prime$}(\textit{$B$}) at different temperatures. A significant intensity difference appears between \textit{$\chi$$^\prime$} and d\textit{M}/d\textit{B} at the boundaries of the field-induced spin textures. In a conventional magnetic system, d\textit{M}/d\textit{B} and \textit{$\chi$$^\prime$}(\textit{$B$}) are expected to exhibit similar peak intensities. However, the discrepancy observed here suggests that the system's response to dynamic magnetic field is delayed at these phase boundaries, indicating the presence of slow relaxation in the system \cite{Bauer2012, Bauer2010, Butykai2017, Bannenberg2018MagnetizationMn1-xFexSi, Qian2016PhaseOSeO3, Bannenberg2016, Gregory1992}.

\subsection{Frequency-dependent \textit{ac} susceptibility \textit{$\chi$$^\prime$}(\textit{$T$})} 
To understand the slow relaxation processes in EuAuBi, we performed temperature-dependent \textit{ac} susceptibility measurements, \textit{$\chi$}(\textit{T}), at different frequencies of the applied \textit{ac} field, superimposed with a specific strength of \textit{dc} field. For \textit{dc} fields, \textit{B} = 0 T and \textit{B} = 1 T, all the \textit{$\chi$$^\prime$}(\textit{T}) for different frequencies overlap within the measured range (Fig 6a and 6b). This suggests the absence of slow magnetic relaxation processes under these specific field conditions or at least within the timescale probed by this frequency range.
As the magnetic field varies, the features of the \textit{$\chi$$^\prime$}(\textit{T}) curves evolve, which are discussed in detail in the following section. 
\newline
To analyze the spin dynamics at higher fields, we recorded frequency-dependent \textit{$\chi$$^\prime$}(\textit{T}) measurements for magnetic field in the range 1.5 T $\leq$ \textit{B} $\leq$ 3 T. Fig 6c and 6d show the results for \textit{B} = 2.5 T and \textit{B} = 2 T, respectively. The peak around 4 K (\textit{$T_{N1}$}), remains at the same position for all excitation frequencies, whereas the lower temperature hump/peak shifts towards higher temperature with increasing frequency. This frequency-dependent \textit{$\chi$$^\prime$}(\textit{T}) behaviour is indicative of a glassy magnetic state, where spins form clusters and exhibit slow magnetic relaxation \cite{Mydosh2014SpinGlasses}.
\newline
To determine the spin relaxation time (\textit{$\tau$$_0$}) in this system, we applied the critical-slowing down model \cite{Mydosh2014SpinGlasses, Manna2011OnCrystals,  Wu2003GlassyLa1xSrxCoO3} which is given by: \textit{$\tau$} = \textit{$\tau$$_0$} $(\textit{T$_f$}/\textit{T$_g$} - 1)^{-z\nu}$, where \textit{$\tau$} is inverse of frequency (\textit{f}), \textit{T$_f$} is the spin freezing temperature at a particular frequency \textit{f}, \textit{T$_g$} is the spin freezing temperature at \textit{f} $\rightarrow$ 0, z\textit{$\nu$} is dynamical critical constant, and \textit{$\tau$$_0$} is the microscopic spin flipping time. The intercept of the curve $log_{10}$\textit{$\tau$} vs $log_{10}$(\textit{T$_f$/T$_g$} - 1) (shown in fig 6f) corresponds to the spin relaxation time $\tau$$_0$ $\sim$ 1.6 × $10^{-7}$ s. This high relaxation time strongly suggests the presence of correlated spin texture in the system rather than individual spin flips in this particular field window \cite{Yang2022CriticalCo7Zn8Mn5, Kindervater2019}.

 \subsection{ \textit{$\chi^\prime$}(\textit{$T$}) and magnetization \textit{$M$}(\textit{$T$}) at different applied fields}
 Temperature-dependent \textit{ac} susceptibility measurements, \textit{$\chi^\prime$}(\textit{T}), were performed under various \textit{dc} magnetic fields superimposed on an \textit{ac} excitation field of 9 Oe and frequency 93 Hz. Distinct features are observed in the \textit{$\chi^\prime$}(\textit{T}) (Fig 7), as \textit{dc} magnetic field varies.
 
  \begin{figure*}[t]
{
    \centering   
    \includegraphics[width=17cm]{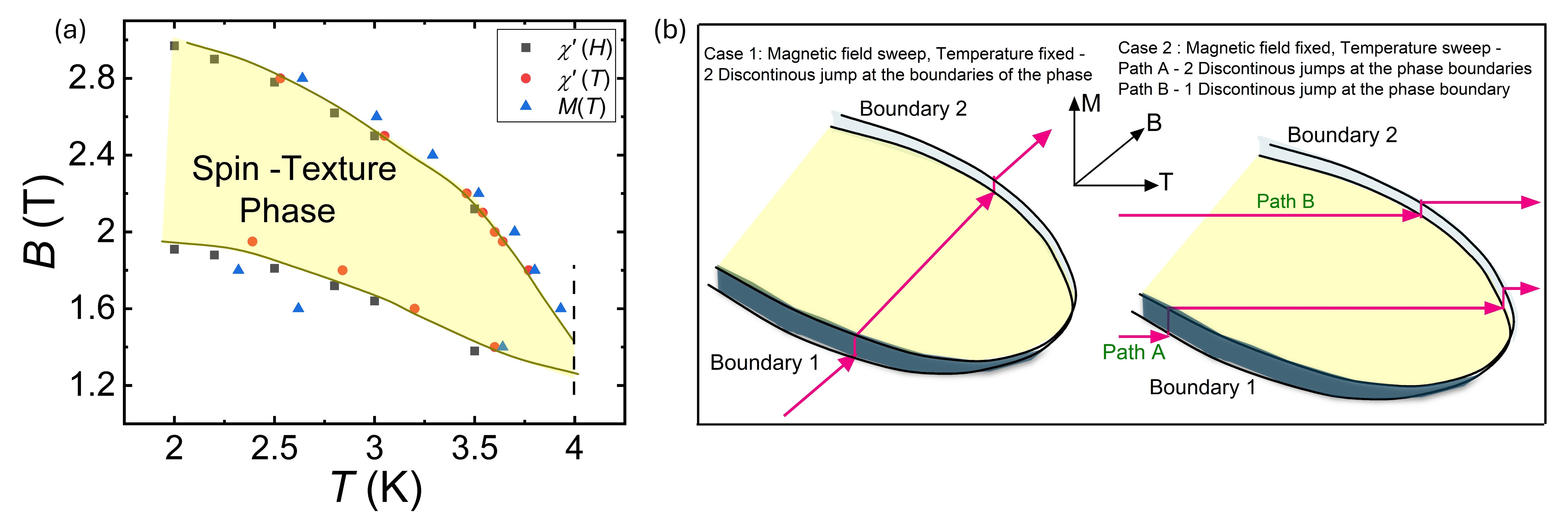}
    \caption{ (a) The non-trivial spin texture phase boundary in EuAuBi is mapped through the \textit{ac} and \textit{dc} magnetization. (b) The schematic shows the jump in magnetization when the boundary of the the field-induced intermediate phase is encountered by varying temperature and magnetic field. }
    \label{fig:my_label9}
}
\end{figure*}
In the absence of a \textit{dc} field, a hump is observed around 3 K, which likely corresponds to some field-induced magnetic spin reorientation (we couldnot identify this transition by zero-field neutron diffraction studies). As the magnetic field increases, this peak broadens due to the emergence of additional features. At 0.05 T (fig 7b), a secondary component starts to appear, which evolves into a more distinct peak at 0.1 T (fig 7c), shifting towards lower temperature with increasing field. This feature is likely associated with the field-induced modulated spin states (possibly helical or conical phase) commonly seen in non-centrosymmetric systems \cite {Levati2014}.
As the field further increases, the relative contribution of the zero-field peak decreases or shifts towards lower temperature, while other components begin to dominate in the \textit{$\chi$$^\prime$}(\textit{$T$}) response. Consequently, the peak around 4 K which corresponds to the Neel temperature also becomes prominent. 
\newline
At 1.4 T (fig 7g), a new anomaly emerges around 3.6 K, marking the onset of possible intermediate field-induced correlated spin texture as previously identified from \textit{$\chi$$^\prime$}(\textit{$B$}) measurements. As the \textit{dc} field increases, this feature becomes more pronounced as it shifts away from the Néel temperature. Interestingly, at 1.8 T, another kink appears at 3.8 K (fig 7i), possibly indicating the second boundary of this phase (as marked by the arrow). At this field, the field-induced correlated phase spans a temperature range from 2.8 K and 3.8 K. Between magnetic fields 1.8 T to 2 T, both temperature boundaries of the this phase remain visible. However, above 2 T, only the second boundary remains as the first boundary shifts below 2 K.
For \textit{dc} fields above 3 T, the second boundary also disappears, likely shifted to lower temperature regime (below 2 K). These observations are summarized in Fig 9, which maps this non-trivial spin texture phase in EuAuBi. 
\begin{figure*}[t]
    \centering   
    \includegraphics[width=17cm]{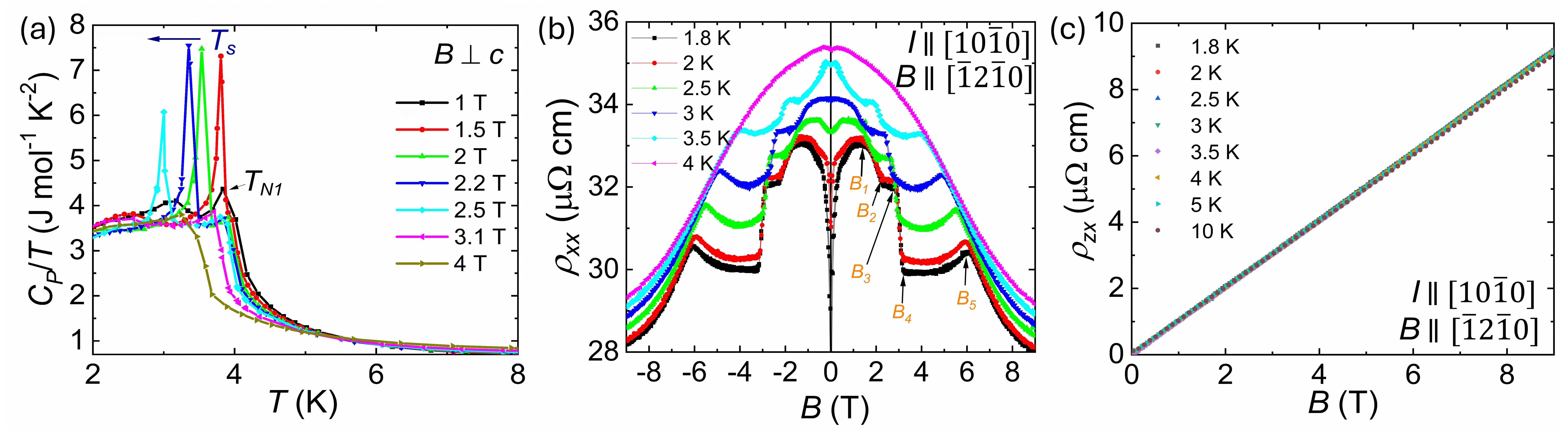}
    \caption{(a) Temperature-dependent specific heat capacity (\textit{C}$_P$(\textit{T})) at different magnetic fields for \textit{B} $\perp$ \textit{c}. \textit{T$_S$} is marked as the first-order phase transition peak temperature (similar to fig 4d), and it shifts towards lower temperature as \textit{B} increases, and \textit{$T_{N1}$} is the Néel temperature. Field dependent (b) longitudinal resistivity \textit{$\rho_{xx}$}(\textit{B}) for B $\parallel$ [$\Bar{1}$2$\Bar{1}$0] and I $\parallel$ [10$\Bar{1}$0] at different temperatures. \textit{B$_1$}, \textit{B$_2$}, \textit{B$_3$}, \textit{B$_4$}, and \textit{B$_5$} marked in the figure are the fields of transition. \textit{B$_1$} - \textit{B$_4$} are also marked in fig 4c. (c) Hall resistivity data \textit{$\rho_{zx}$}(\textit{B}) at different temperatures for \textit{B} $\parallel$ [$\Bar{1}$2$\Bar{1}$0] and \textit{I} $\parallel$ [10$\Bar{1}$0].}
    \label{fig:my_label10}
\end{figure*}
\newline
To further confirm the evolution of magnetic phases, we correlated the features of \textit{$\chi$$^\prime$}(\textit{$T$}) with the \textit{M}(\textit{T}) measurements in similar field regions. At the boundaries of the non-trivial spin texture phase as identified via \textit{$\chi$$^\prime$}(\textit{$T$}) measurements (fig 7g - 7n), \textit{M}(\textit{T}) is expected to show sudden jumps. As shown in Figure 8, the \textit{M}(\textit{T}) exhibit features consistent with the \textit{$\chi$$^\prime$}(\textit{$T$}) data. At 1.4 T, a single jump is observed in the \textit{M}(\textit{T}) curve at 3.6 K (fig 8a), matching the onset of one boundary of this magnetic phase as observed in \textit{$\chi$$^\prime$}(\textit{$T$}). Between 1.8 T and 2.1 T, two distinct jumps appears in \textit{M}(\textit{T}), marking both phase boundaries of the non-trivial spin texture phase. At higher fields, only one jump remains, as the first boundary shifts to lower temperatures and eventually falls below 2 K. This consistency between the \textit{M}(\textit{T}) and \textit{$\chi$$^\prime$}(\textit{$T$}) measurements provides a coherent picture of the field- and temperature-boundaries of the possible correlated non-trivial magnetic phase in EuAuBi.
However, the \textit{ac} susceptibility study provides only indirect evidence, indicating the need for further investigations to confirm the nature of this field-induced non-trivial magnetic phase in the EuAuBi system.
\newline
Based on our detailed analysis of \textit{$\chi$$^\prime$}(\textit{$T$}), \textit{$\chi$$^\prime$}(\textit{$B$}), \textit{M}(T), and \textit{M}(\textit{B}) measurements, we construct the magnetic phase boundary of this intermediate phase in EuAuBi (Fig 9a). This non-trivial magnetic phase exhibits a characteristic tilted plateau in the magnetization profile as shown in fig 9b -
\newline
Case 1: Magnetic field sweep at fixed temperature -
When the magnetic field is swept at a constant temperature, two distinct jumps are observed in the magnetization data. These jumps correspond to the system entering and exiting the intermediate non-trivial phase, consistent with the proposed phase diagram.
\newline
Case 2: Temperature sweep at fixed magnetic field - 
When the temperature is swept at a constant magnetic field, the system shows two possible scenarios depending on its trajectory through the phase diagram.
\newline
• Path A: Both boundaries of the intermediate phase are intercepted, resulting in two jumps in the \textit{M}(\textit{T}). This behavior is observed in fields above 1.6 T and up to 2 T.
\newline
• Path B: Only one boundary of the intermediate phase is intercepted, leading to a single jump in the \textit{dc} \textit{M}(\textit{T}).
\newline
 The similarity of this intermediate field-induced correlated spin-texture to behaviors observed in established skyrmion-hosting systems is noteworthy \cite{Bauer2012,Bauer2010,Butykai2017,Qian2016PhaseOSeO3, Bannenberg2018MagnetizationMn1-xFexSi, Gregory1992}. However, considering that \textit{ac} studies provide only indirect evidence, direct imaging techniques are essential to unambiguously identify the nature of this intermediate state.

\subsection{Specific heat and electrical transport measurements}
To investigate the nature of the intermediate field-induced correlated spin texture, we performed temperature-dependent specific heat capacity measurements, \textit{C}$_P$(\textit{T}), at different magnetic fields. The commensurate antiferromagnetic transition at \textit{$T_{N1}$} appears as \textit{$\lambda$}-type peak (indicative of second-order phase transition) around 4 K in all \textit{C}$_P$(\textit{T}) curves.
\newline
At B = 1 T, a hump-like feature appears around 3 K, likely associated with the spin-modulated state (possibly helical/conical phase), consistent with the \textit{$\chi$$^\prime$}(\textit{$T$}) results. With further increase in the \textit{dc} field to 1.5 T, a prominent \textit{$\delta$}-type peak emerges around 4 K, overlapping with the Néel temperature \textit{$\lambda$}-peak. The \textit{$\delta$}-type peak is a characteristic of a first-order transition, coinciding with the onset of the non-trivial spin texture phase as previously identified in the magnetization measurements. 
With increasing magnetic field, the hump associated with the spin modulated phase shifts to lower temperatures and eventually vanishes above 2 T. Simultaneously, the \textit{$\delta$}-type also shifts towards the lower temperature and gradually decreases in intensity until it disappears around 3 T. This implies the suppression in magnetic phase transition with increase in magnetic field. Hence, above the intermediate phase regime, only the \textit{$\lambda$}-type peak corresponding to antiferromagnetic–paramagnetic transition (\textit{$\lambda$}-peak) persists. These observations provide additional evidence for a first-order phase transition associated with the non-trivial spin texture phase in the EuAuBi system \cite{Bauer2013,Mishra2021}.
\newline
To further investigate the impact of field-induced complex spin states on charge transport, we conducted electrical transport measurements. Figure 10b illustrates the field-dependent resistivity, \textit{$\rho_{xx}$}(\textit{B}), at different temperatures for \textit{B} $\parallel$ [$\Bar{1}$2$\Bar{1}$0] and \textit{I} $\parallel$ [10$\Bar{1}$0]. Above the antiferromagnetic transition around 4 K, resistivity decreases with increasing magnetic field, attributed to reduced spin-dependent scattering within the system. However, the crystals exhibit distinct transport behavior below \textit{T$_{N1}$} ($\sim$ 4 K). At low magnetic fields (up to 0.3 T), a decrease in resistivity is observed across all isotherms below 2.5 K. A previous study \cite{Takahashi2023} attributed this behavior to intrinsic superconductivity in EuAuBi. However, the absence of both: the superconducting transition in \textit{$\rho_{xx}$} data and the Meissner effect in \textit{M}(\textit{T}) experiments in our EuAuBi crystals suggests that this resistivity drop is more likely due to extrinsic effects, possibly arising from impurity phases or defects in the system, potentially caused by residual Bi flux or Au–Bi binary inclusions formed during the crystal growth process.
\newline
As the magnetic field increases, \textit{$\rho_{xx}$} first increases and reaches a maximum around \textit{B$_1$}, then it sharply drops until \textit{B$_2$}, and remains constant up to \textit{B$_3$}, followed by another steep decrease in \textit{$\rho_{xx}$} until \textit{B$_4$}. This anomaly region from \textit{B$_1$} to \textit{B$_4$} exactly matches with the intermediate non-trivial spin texture phase regime as observed in the magnetization measurements shown in Figs 4c. This correspondence suggests that the interaction between the conduction electrons and the spin-texture results in reduced scattering, and hence a drop in resistivity \textit{$\rho_{xx}$}(\textit{B}). Furthermore, above 2 K, the decrease in the resistivity curve between \textit{B$_1$} and \textit{B$_2$} becomes less distinct, and the \textit{B$_3$} - \textit{B$_4$} shifts towards lower magnetic fields. Similar behavior is observed in \textit{M}(\textit{B}) curves for \textit{B} $\perp$ \textit{c} (Fig 4c) also. Between \textit{B$_4$} and \textit{B$_5$}, system maintains the antiferromagnetic state and \textit{$\rho_{xx}$} is nearly flat. However above \textit{B$_5$}, \textit{$\rho_{xx}$} decreases due to lower spin-dependent scattering in the system. 
\newline
\begin{figure}[t]
    \centering   
    \includegraphics[width=\columnwidth]{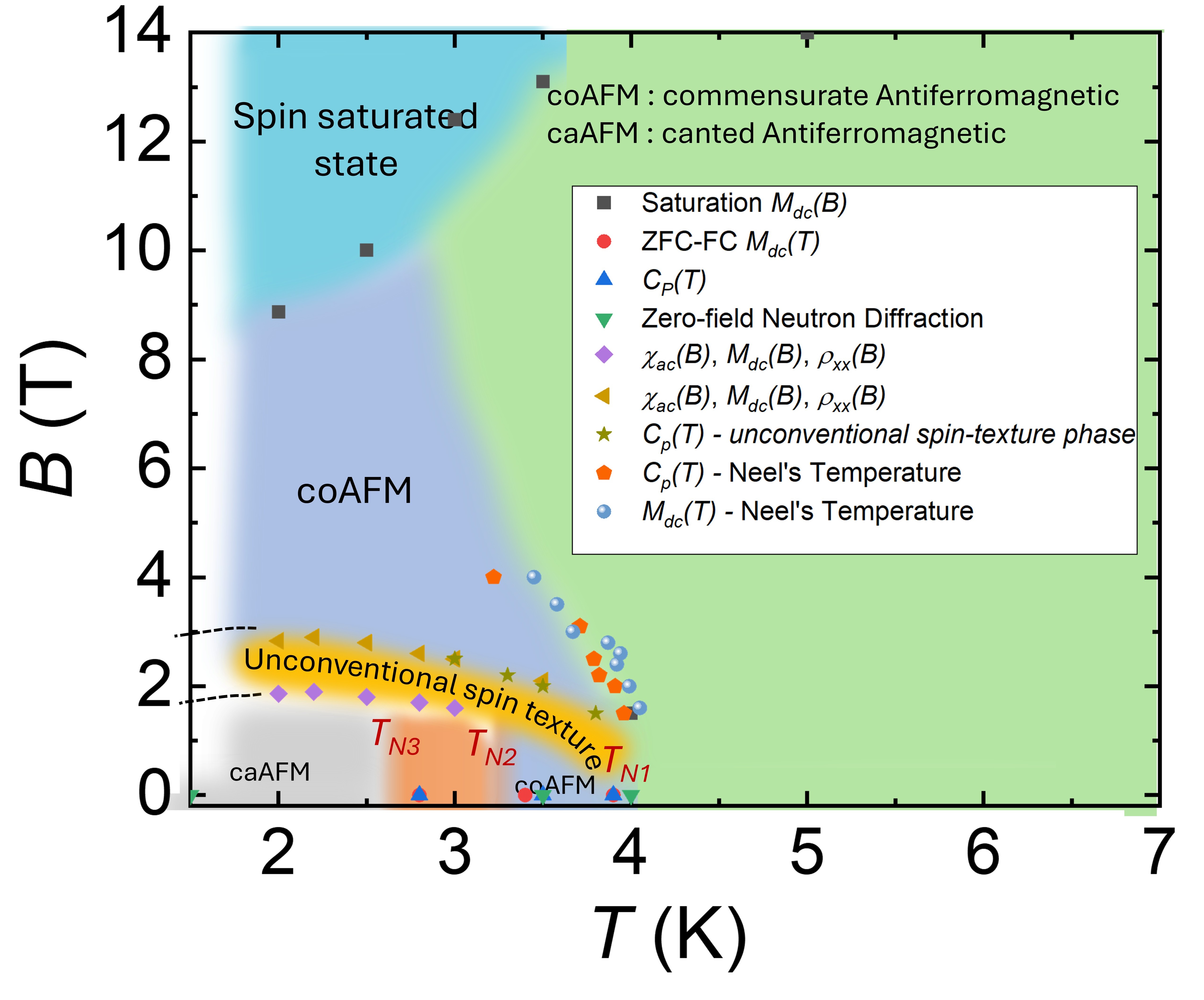}
    \caption{Complex magnetic phase diagram of the EuAuBi system for \textit{B} $\perp$ \textit{c}
     derived from detailed neutron diffraction, magnetization, specific heat and electrical transport experiments. caAFM denotes canted antiferromagnetic phase and coAFM denotes commensurate antiferromagnetic state. The phase between \textit{T$_{N2}$} and \textit{T$_{N3}$} is unknown phase as it could not be resolved using neutron diffraction.}
    \label{fig:my_label11}
\end{figure}
Fig 10c shows the Hall resistivity data ($\textit{$\rho_{zx}$}$) at different temperatures with \textit{B} $\parallel$ [$\Bar{1}$2$\Bar{1}$0], \textit{I} $\parallel$ [10$\Bar{1}$0], and Hall voltage,  \textit{V} $\parallel$ [0001]. The positive slope suggests hole-dominant charge carriers in the system. The charge carrier concentration and their mobility in the system are calculated from \textit{$\rho_{zx}$}(\textit{B}) as $\sim$ 6.11 X $10^{20}$ /cm$^3$ and 189.3 cm$^2$/V-s respectively at 2 K. Besides, the carrier concentration is observed almost temperature independent upto 10 K, as the slope of \textit{$\rho_{zx}$}(\textit{B}) is nearly same for all the temperatures. It is important to note that the present Hall measurement configuration is defined by the specified directions of the applied magnetic field and current. Under this configuration, the system does not exhibit any topological Hall effect, in contrast to other skyrmion-like systems. However, it has been observed that the topological Hall effect is highly sensitive to the measurement configuration. Consequently, Hall resistivity should also be measured along other orientations such as \textit{B} along [10$\Bar{1}$0] and \textit{I} along [001]. However, the Hall experiments were unsuccessful along these orientations due to the thin, plate-like morphology of the grown crystal, which restricts the electrical contacts alignment along these directions.
\newline
Following the comprehensive investigation of the neutron diffraction, magnetic, specific heat and electrical transport measurements of the EuAuBi crystal, we have constructed the experimental magnetic phase diagram for \textit{B} $\perp$ \textit{c} (Fig 11). It summarizes the coexistence of different magnetic phases and complex field-induced transitions in the EuAuBi system. 
\newline

\section{Conclusion} 
In summary, we have presented a comprehensive investigation of the complex spin dynamics in single crystals of EuAuBi using neutron diffraction, magnetization, electrical transport, and specific heat measurements. Density functional theory calculations reveal symmetry-protected Dirac band crossing close to the Fermi energy (E$_F$) in EuAuBi. The magnetic and specific heat measurements reveal three distinct temperature-dependent transitions below $\sim$4 K, and zero-field-powdered neutron diffraction identifies two of these magnetic spin arrangements as a commensurate antiferromagnetic phase at 4 K, and a canted antiferromagnetic phase at 1.5 K. The intermediate transition could not be resolved unambiguously due to the high neutron absorption of europium. Field- and temperature-dependent \textit{dc} magnetization measurements demonstrate pronounced anisotropy, with a clear field-induced phase transition observed for \textit{B} $\perp$ \textit{c} within a specific field and temperature window. Consequently, the frequency-, field- and temperature-dependent features in \textit{ac} susceptibility indicates the possible presence of topologically protected, non-trivial spin textures with slow relaxation times in the intermediate field regime. Notably, EuAuBi exhibits a tilted plateau-like region in the magnetization, resembling the skyrmion hosting systems. The interaction between conduction electrons and these spin textures is also evident in the magnetoresistivity anomalies. All the experimental results consistently indicate the possible presence of a non-trivial spin texture in the EuAuBi. Based on these observations, we have constructed a comprehensive magnetic phase diagram that captures the evolution of field-induced spin texture in the system. Our findings suggest that EuAuBi hosts a complex spin-texture-induced topological phase, making it a model system in which Berry curvature emerges in both real and momentum space.
\begin{acknowledgments}
We acknowledge Max Planck
Society for the funding support under Max Plank-India partner group project, and Board of Research in Nuclear Sciences (BRNS) under 58/20/03/2021-BRNS/37084/ DAE-YSRA, Science and Engineering Research Board, DST, Government of India, via grant no: CRG/2022/001826 and
Aeronautics Research and Development Board (ARDB, Project No. 1992). Lipika would like to thank the Prime Minister Research fellowship grant (PMRF Id : 1402122)  for the research support and fellowship. Authors also thanks Central research facility (CRF) IIT Delhi and Nanoscale research facility at IIT Delhi for providing for providing the materials characterization facility, like PPMS, MPMS, EDX, XRD etc.
\end{acknowledgments} 
\bibliography{bibliography.bib}

\end{document}